\documentclass[prb,twocolumn,showpacs,preprintnumbers]{revtex4}
\usepackage{graphics}
\usepackage{graphicx}
\usepackage{dcolumn}
\usepackage{bm}
\usepackage{float}
\usepackage{epsfig}
\begin{document}
\title{Cohesion of BaReH$_9$ and BaMnH$_9$: Density Functional Calculations
and Prediction of (MnH$_9)^{2-}$ Salts}
\author{D.J. Singh}
\address{Materials Science and Technology Division, Oak Ridge National
Laboratory, Oak Ridge, TN 37831-6032}
\author{M. Gupta}
\address{
Thermodynamique et Physico-Chimie d'Hydrures et Oxydes, EA3547, Batiment 415, 
 Science des Materiaux,
Universite Paris-Sud, 91405 Orsay, France}
\author{R. Gupta}
\address{Service de Recherches de Metallurgie Physique,
Commissariat a l'Energie Atomique, Centre d'Etudes de Saclay, 
91191 Gif Sur Yvette Cedex, France}
\date{\today}

\begin{abstract}
Density functional calculations are used to calculate the
structural and electronic properties of BaReH$_9$ and to
analyze the bonding in this compound.
The high coordination in
BaReH$_9$ is due to bonding between Re 5$d$ states
and states of $d$-like symmetry formed from combinations
of H $s$ orbitals in the H$_9$ cage. This explains the structure of the
material, its short bond lengths and other physical properties, such
as the high band gap.
We compare with results for hypothetical
BaMnH$_9$, which we find to have similar bonding
and cohesion to the Re compound. This suggests that it may
be possible to synthesize (MnH$_9)^{2-}$ salts.
Depending on the particular cation, such salts may have exceptionally
high hydrogen contents, in excess of 10 weight \%.
\end{abstract}


\maketitle

\section{Introduction}

BaReH$_9$ is the prototypical member of a family of hydrides, based
on (ReH$_9)^{2-}$ and (TcH$_9)^{2-}$ structural units.
\cite{knox,gin1,ginsberg,stetson,stetson2,ryan}
These compounds are of interest from a fundamental point
of view
\cite{pauling}
because of the unusual coordination of the transition metal
atoms, which are surrounded by 9 hydrogen atoms, with relatively
short metal-H bond lengths and are in an high formal
valence state of 7.
These compounds are also of practical interest because
of the high hydrogen to metal ratio of 4.5 in BaReH$_9$.
Although no Mn based examples of these compounds have been
synthesized to date, the hypothetical 3$d$ analogue, MgMnH$_9$,
would have a hydrogen content in excess of 10 weight percent.
The purpose of this paper is to analyze the electronic structure of
BaReH$_9$ in order to understand its bonding and the prospects
for synthesis of Mn based analogues.
This follows a previous electronic structure calculation
by Orgaz and Gupta, \cite{orgaz} done using the X-ray crystal
structure of Ref. \onlinecite{stetson} and
the $X\alpha$ method. The present calculations were
done using a general potential self-consistent method, with
calculated H positions, and allow us to present a more
accurate electronic structure and detailed analysis of the bonding.

\begin{figure}
\vskip 0.4 cm
\includegraphics[width=3.0in]{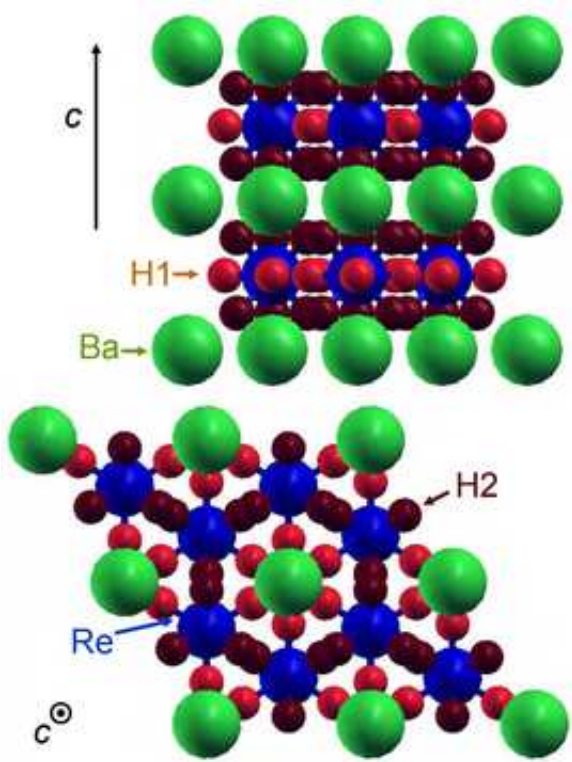}
\vskip 0.4 cm
\caption{(color online) Structure of BaReH$_9$, showing Ba and ReH$_9$
layers stacked along the $c$-axis (top) and the structure of
the hexagonal planes (bottom). The H1 and H2 sites are shown
by small light and dark red spheres, respectively, Ba by large
light green spheres, and Re by large dark blue spheres. The
atomic coordinates are those obtained by LDA structure
relaxation.}
\label{structure} \end{figure}

The crystal structure (spacegroup $P6_3/mmc$), which was
determined using X-ray diffraction by Stetson and
co-workers, \cite{stetson}
is depicted in Fig. \ref{structure}. It
consists of alternating triangular layers of Ba and Re stacked along the
$c$-axis, such that the Re are at the center of trigonal prisms
formed by the Ba. Each Re is coordinated by three H atoms
in its own plane (the H1) positions, and six other H atoms (the H2)
above and below the plane. These H2 atoms form a trigonal prism. Thus the
H atoms are arranged in tricapped trigonal prisms around the Re atoms.
The unit cell contains two (ReH$_9)^{2-}$ units, stacked so that
the orientation of these units alternates along the hexagonal $c$-axis.

\section{Method}

The calculations reported here were done within the local density 
approximation (LDA) using the general potential linearized augmented
planewave (LAPW) method. \cite{singh-book}
Local orbitals \cite{singh-lo}
were used for the semicore states (Re 5$s$ and 5$p$,
and Ba 5$s$ and 5$p$), and to relax the linearization of the $d$
bands. LAPW sphere radii of 2.0 $a_0$ and 1.1 $a_0$ were
used for the metal and H atoms, respectively.
We used well converged basis sets consisting of more than 3200
LAPW functions and local orbitals for the two formula unit
primitive cell. We tested larger basis sets, but found
no significant changes in the results.
We also tested various Brillouin zone samplings for the iteration
to self-consistency, but found that the results were already converged
at a sampling of 6 special {\bf k}-points in the irreducible wedge
of the hexagonal zone. No doubt this reflects the large band
gap insulating character of the compound.
The calculations for hypothetical BaMnH$_9$ were done
with similar parameters, except that the Mn and H radii were
1.8 $a_0$ and 1.05 $a_0$, respectively.

\section{Structure}

While the lattice parameters, metal positions,
and basic structure determined by
Stetson and co-workers (Ref. \onlinecite{stetson})
are undoubtedly correct,
it is difficult
to accurately determine H positions in a complex compound containing
heavy atoms without neutron scattering. Accordingly, we used the
experimental lattice parameters ($a$=5.287\AA, $c$=9.323\AA) here,
but determined the internal
coordinates of the H atoms by energy minimization within the LDA.
The calculated structural parameters are given in Table \ref{struct-tab}
along with those of hypothetical BaMnH$_9$, which is discussed
later.
As may be seen, the tricapped trigonal prisms are quite compact and
regular in
that the Re-H1 and Re-H2 bond lengths are short and similar, although
there is a compression along the $c$-axis, as may be seen from the
fact the the H2-H2 neighbor distance on the top of the trigonal
prism is longer than the H1-H2 distance between the prism and
cap hydrogen sites.
It is also notable that the H1-H2 distance
of 1.92\AA$~$ is very short compared with most metal hydrides, which
generally have H - H distances larger than 2.1 \AA. \cite{westlake1,westlake2}
These short H-H distances within the (ReH$_9)^{2-}$ units
suggest that direct H-H interactions may be important in forming
the electronic structure, which in fact is what we find.

\begin{table}[tbp]
\caption{Calculated structural parameters of  $P6_3/mmc$ BaReH$_9$
and hypothetical BaMnH$_9$, obtained
within the LDA. The lattice parameters were fixed at the
values $a$=5.287\AA, and $c$=9.323\AA, as reported
in Ref. \onlinecite{stetson} for BaReH$_9$.
For hypothetical BaMnH$_9$ we used $a$=5.067\AA, and $c$=8.883\AA$~$
(see text).
The coordinates of the
atoms are Ba (2$a$): (0,0,0), Re/Mn (2$c$): (1/3,2/3,1/4),
H1 (6$h$): ($x_1$,2$x_1$,1/4) and H2
(12$k$): ($x_2$,2$x_2$,$z_2$). Distances, $d$, are
the shortest bonds of a given type.}
\begin{tabular}{lcc}
\hline
parameter~~~~~ & ~~BaReH$_9$~~ &~~ BaMnH$_9$~~ \\
\hline
$x_1$ & 0.151 & 0.161 \\
$x_2$ & 0.466 & 0.460 \\
$z_2$ & 0.122 & 0.129 \\
$d$(Re/Mn-H1)   & 1.67\AA & 1.51\AA \\
$d$(Re/Mn-H2)   & 1.71\AA & 1.54\AA \\
$d$(H1-H2)      & 1.92\AA & 1.73\AA \\
$d$(H2-H2)      & 2.10\AA & 1.92\AA \\
$d$(H1-H1)$^*$  & 2.39\AA & 2.44\AA \\
$d$(Ba-H1)      & 2.71\AA & 2.63\AA \\
$d$(Ba-H2)      & 2.89\AA & 2.80\AA \\
\hline
$^*$between different (ReH$_9)^{2-}$ units.
\end{tabular}
\label{struct-tab}
\end{table}

\begin{figure}
\vskip 0.4 cm
\includegraphics[width=2.0in,angle=270]{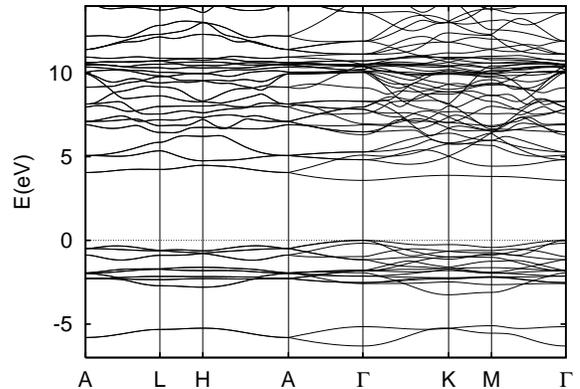}
\vskip 0.4 cm
\caption{Calculated band structure of hexagonal BaReH$_9$,
using the relaxed crystal structure.}
\label{bands} \end{figure}

\begin{figure}
\vskip 0.4 cm
\includegraphics[width=2.4in,angle=270]{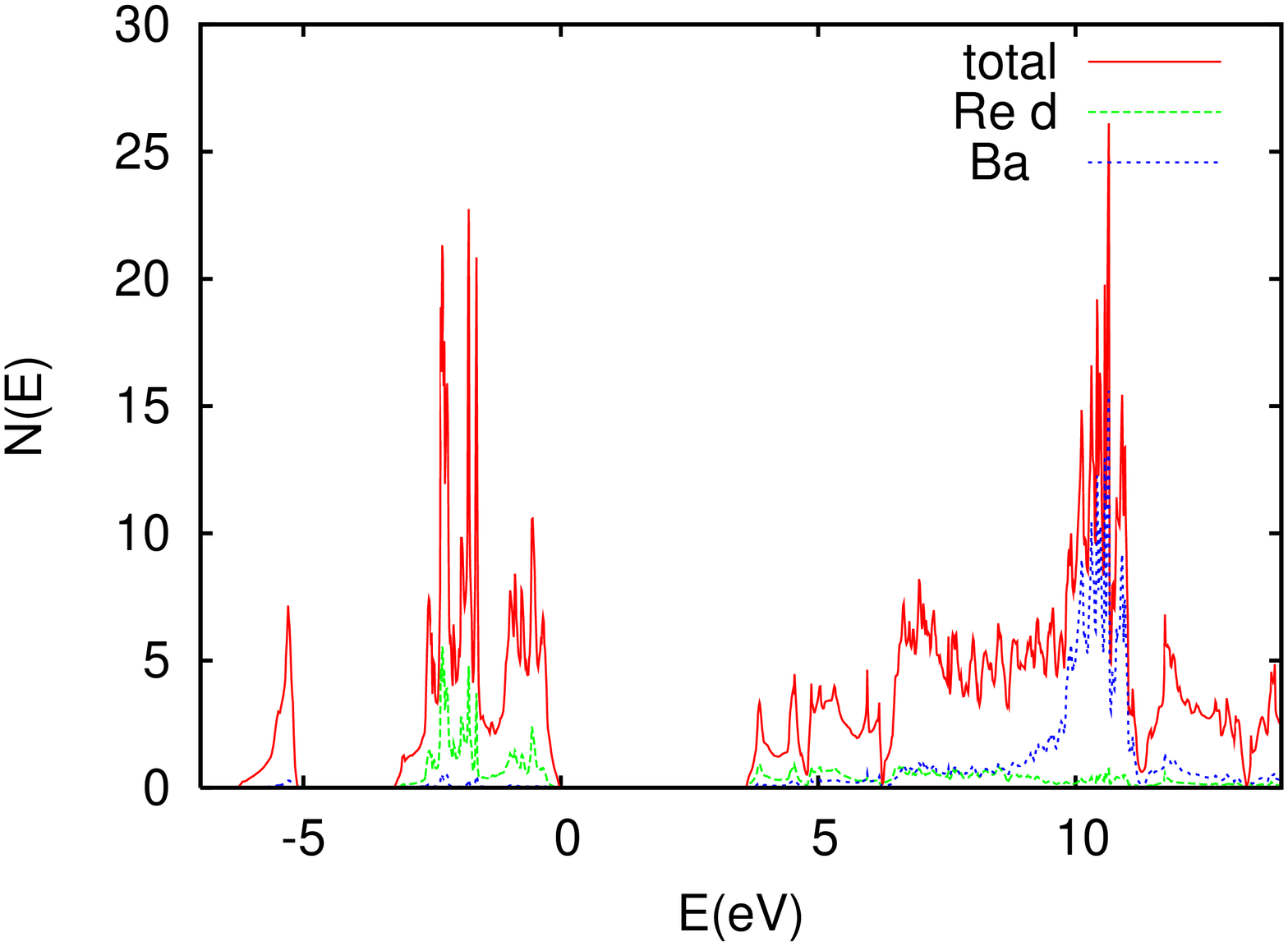}
\includegraphics[width=2.4in,angle=270]{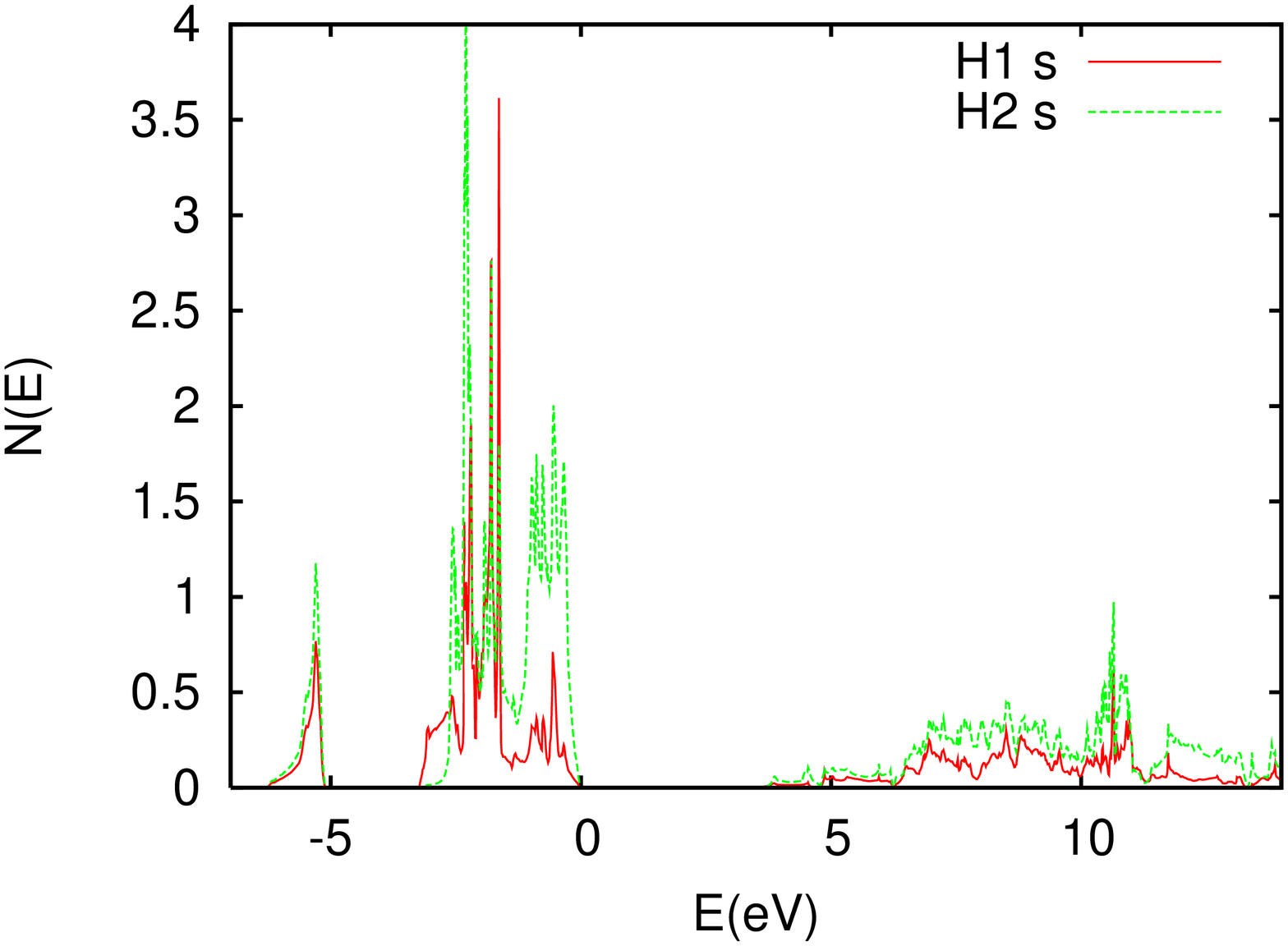}
\vskip 0.4 cm
\caption{(color online) Calculated electronic density of states
(DOS) of BaReH$_9$ on a per formula unit basis.
The top panel shows the total, Re $d$ and Ba contributions, as defined
by projections onto the LAPW spheres.
The bottom panel shows the projections of H1 $s$ and H2 $s$ character.
Note that for 1.1 $a_0$ H LAPW spheres approximately 2/3 of the
H $s$ charge for a H atom would lie outside the sphere.
}
\label{dos} \end{figure}

\section{Electronic Structure}

\begin{figure}
\vskip 0.4 cm
\includegraphics[width=2.4in,angle=270]{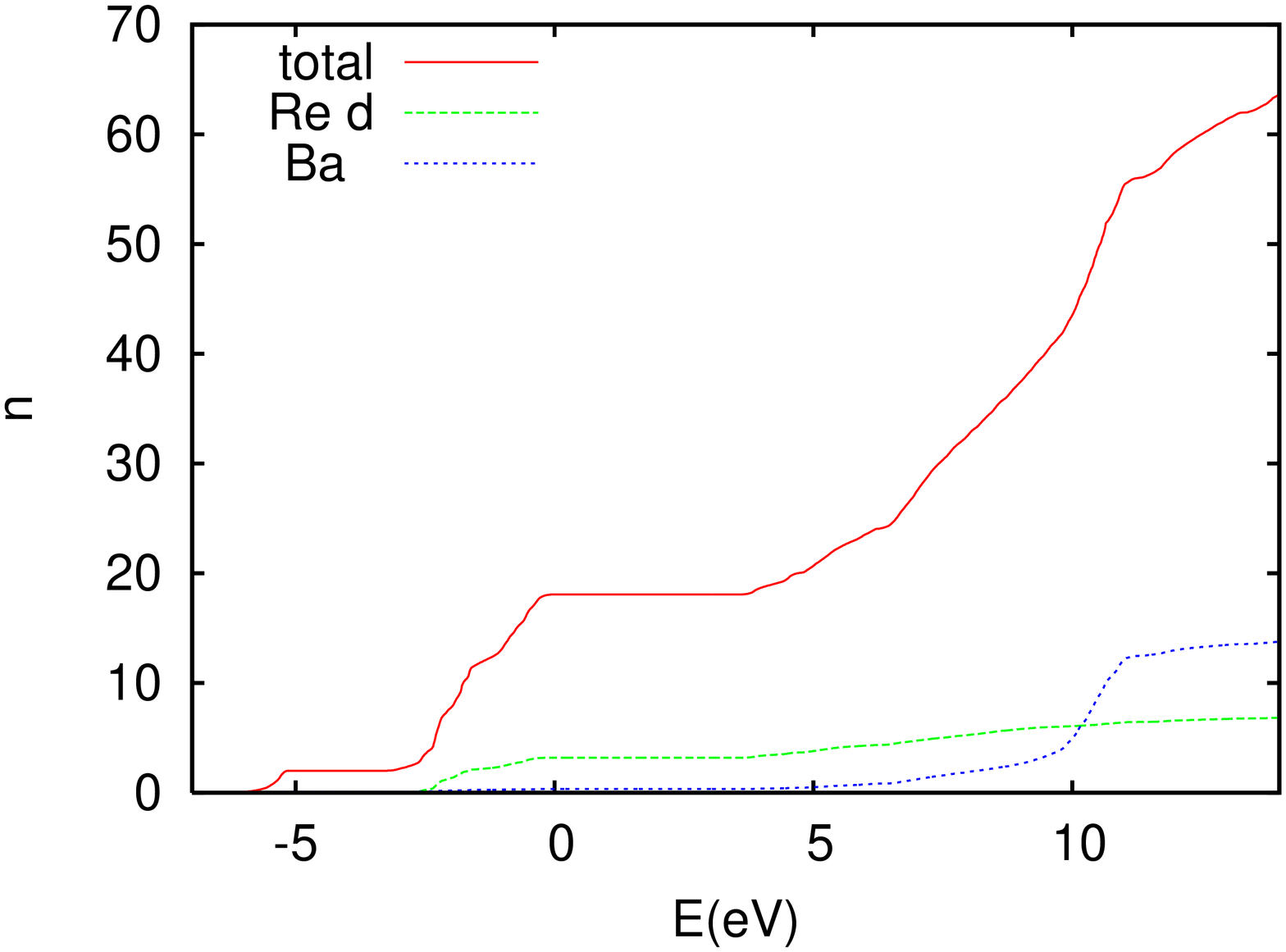}
\includegraphics[width=2.4in,angle=270]{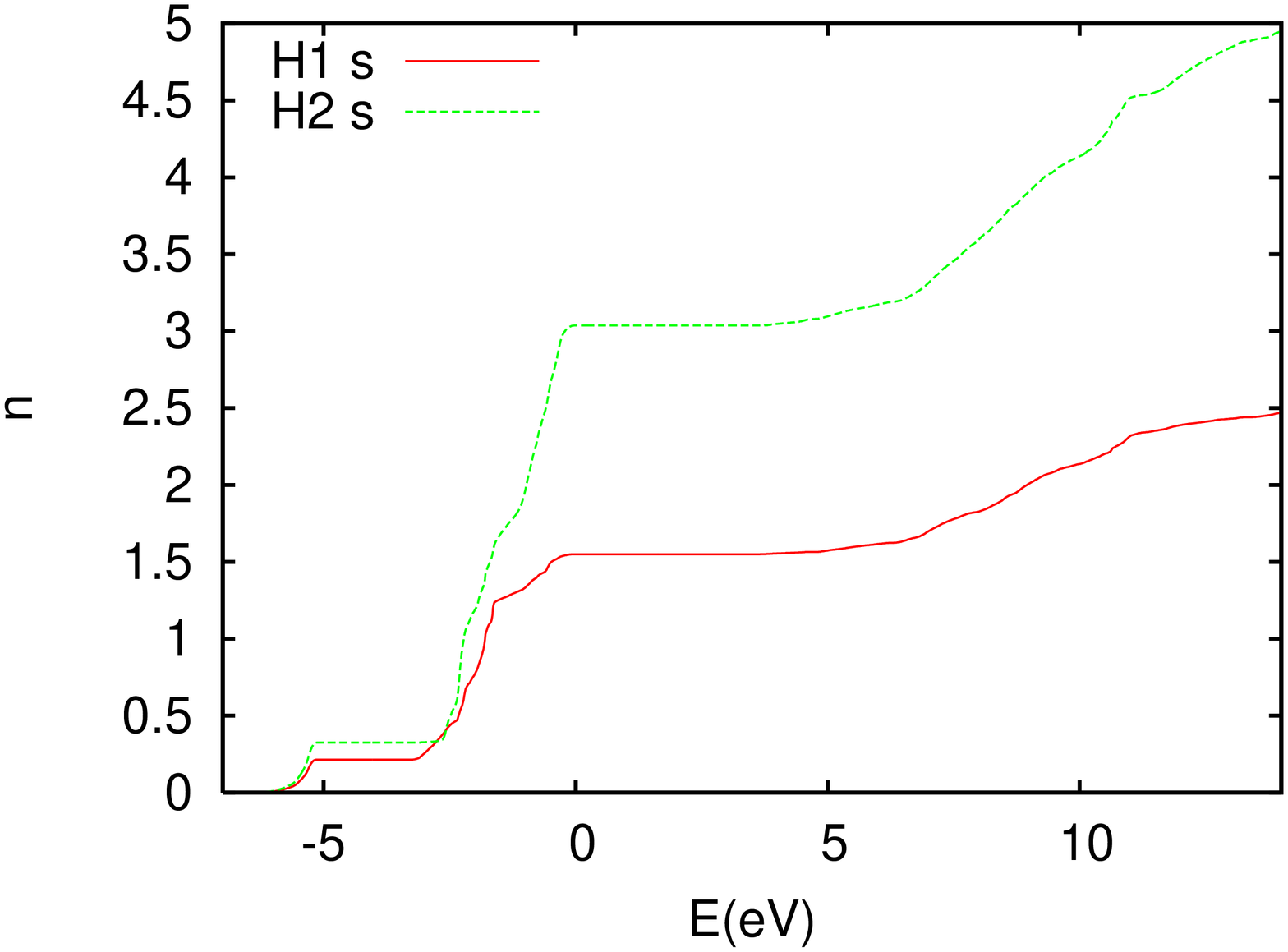}
\vskip 0.4 cm
\caption{(color online) Integrated
DOS of BaReH$_9$ on a per formula unit basis.
The top panel shows the total, Re $d$ and Ba contributions, as in Fig.
\ref{dos}.
The bottom panel shows the projections of H1 $s$ and H2 $s$ character.
The integrations are on a per formula unit basis.
}
\label{integrated} \end{figure}

Our main electronic structure results are given in
Figs. \ref{bands} - \ref{integrated}.
The calculated band structure and electronic density of states (DOS)
are shown in Figs. \ref{bands} and \ref{dos}, respectively.
Fig. \ref{dos} also shows projections of the DOS onto the 
LAPW spheres. Electron counts corresponding to integration
of the DOS and projections are shown in Fig. \ref{integrated}.
As in the calculation of Ref. \onlinecite{orgaz}, we obtain a wide
band gap insulator, but the bands and the structures and positions
of the features in the DOS are significantly different.
We obtain a direct band gap of 3.58 eV at $\Gamma$, which may be
an underestimate, as is common in LDA calculations.
In any case, this large band gap is consistent with
the observed transparent nature of the material, and also
indicates the nature of the band formation. In particular,
it is an extremely large value for a crystal field gap, and also
is not consistent with an ionic gap, since in a scenario
with H$^-$ and Re$^{7+}$ ions, one would expect low
lying Re $d$ levels and therefore a small gap
due to the high Re valence. Such an ionic scenario is also unlikely
considering the short H-H bond lengths in the (ReH$_9)^{2-}$ units.

Examining the band structure in greater detail, we note that there
are three manifolds of states. These are a narrow set of two bands
(one band per formula unit) extending from -6.2 eV to -5 eV (relative
to the valence band maximum), a manifold of 16 bands extending
from -3.3 eV to the valence band maximum and a broad set of conduction
bands. While valence bands are often less dispersive than
conduction bands in semiconductors, this electron-hole asymmetry is
particularly noticeable in the BaReH$_9$. As may be seen from
the projections, of the DOS there is no appreciable Ba
contribution to the valence bands, and therefore, as expected, Ba
occurs as Ba$^{2+}$. There is, however, Ba $d$ character in the upper
conduction bands starting at $\sim$ 7 eV, and the Ba $f$ resonance is visible
from $\sim$ 10 - 11 eV.

With the exception of the lowest, split-off peak in the DOS, which 
is H $s$ in character, the remainder of the DOS shows a mixture
of Re $d$ and H $s$ character, demonstrating covalency between Re and H.
This is distinct from what is seen in NaAlH$_4$, for example.
\cite{aguayo}

The projected densities of states and electron counts are based
on integration within the LAPW spheres. In the LAPW method these
spheres are constrained to be non-overlapping, and in addition
they must be chosen so that core states are contained within them.
These constraints require the use of rather small spheres at the
H sites. Here we used $r_{\rm H}$ = 1.1 $a_0$. This is significant
for the interpretation of the projected DOS, because most of the
H $s$ charge lies outside this radius. In particular,
for a neutral non-spin-polarized H atom in the LDA only 0.329 electrons
are contained within 1.1 $a_0$,
and 0.305 electrons within 1.05 $a_0$. This factor needs to be kept in
mind when analyzing the DOS. The advantage of the small H LAPW sphere
is that the H wavefunction near the nucleus is much less sensitive to
the environment, which means that charge within this size sphere
is almost entirely due to the H $s$ states, with very little
contamination by charge entering from neighboring atoms. We
verified this by integrating the $p$ projection in the H spheres,
which would come from tails of neighbors,
and found less than 0.02 electrons, supporting this conclusion.
Thus, integration of the H $s$ projection can provide a useful 
measure of the H valence, when normalized by the charge of a neutral H.
The charge obtained in this way corresponds to H$^{0.5-}$ for
both the H1 and H2 sites. This analysis cannot be easily applied
to the conduction bands because there is no  plateau at the top of
the Re $d$ derived DOS, presumably because of contributions from the
free electron like states, which would provide H $2s$ and other contributions.
Nonetheless, by this measure, 2 electrons per H would be included
by integrating up to 9 eV, while integrating to 10 eV, where
the Re $d$ character becomes small, yields 2.1 electrons.
In any case, these numbers indicate substantial covalency between
Re and H. This is also consistent with the large bandgap, which
comes from the bonding - antibonding splitting associated with the
covalency.

\section{Bonding}

To understand the bonding, we examined the symmetry of the H $s$ states
when expanded about the Re site, using the projections of the H tails
into the Re sphere. Based on this, the lowest split-off peak
containing 2 electrons comes
from the symmetric combination of the H $s$ states, and is $s$-like
at the Re site. The second peak, which contains 16 electrons per
(ReH$_9)^{2-}$ unit has both $p$-like and $d$-like combinations of
the H $s$ states. Considering the short H-H distance in the H$_9$
cage, and the simple electronic structure of H, the electronic
states of a H shell should be ordered according to the number of nodes,
yielding, in order, a fully symmetric state ($s$-like), three states with
one node ($p$-like) and five states with two nodes ($d$-like).
In fact, calculations for free Be$_9$ and Mg$_9$ clusters,
\cite{kawai,kumar}
which
have 18 electrons and single $s$ orbital ions, show that these
clusters are exceptionally stable and have a tricapped trigonal prismatic
ground state structure, similar to 18 electron (ReH$_9)^{2-}$.
Furthermore, the electronic structure \cite{kumar}
shows clearly separated groups of states, which are, in order,
s-like, p-like and d-like. The energy splittings of the states
within the groups due to the non-spherical, $D_{3h}$, symmetry
are small compared to the splittings between the groups.
This strong shell structure of the orbitals is associated with
the stability of these clusters and their shape, as discussed
by Reimmann and co-workers, \cite{reimann} and may be expected
to be a general tendency in 18 electron, 9 site clusters with only
$s$-orbitals.

\begin{figure}
\vskip 0.4 cm
\includegraphics[width=3.5in,angle=0]{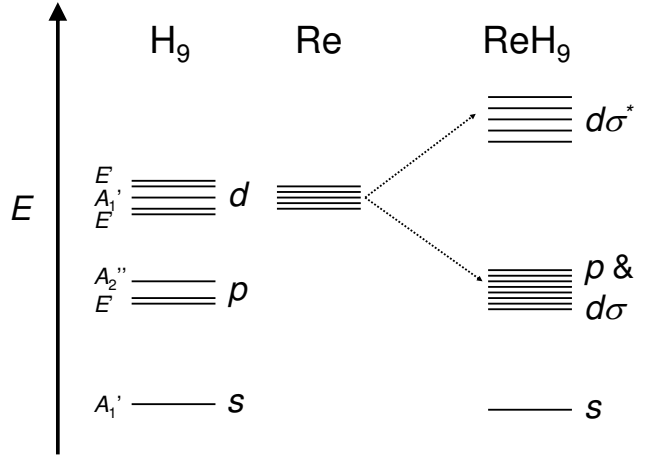}
\vskip 0.4 cm
\caption{Schematic depiction of the formation of the BaReH$_9$
electronic structure from the levels of a tricapped trigonal
prismatic H$_9$ cluster (see Ref. \onlinecite{kumar})
and
the Re $d$ orbitals. The levels of the cluster are labeled according
using $D_{3h}$ representations, and also $s$, $p$ and $d$ according to
their approximate angular momentum character when expanded around
the center of the cluster. Note that all the H states shown are derived from
different linear combinations of H 1$s$ orbitals.
}
\label{levels} \end{figure}

This forms the basis for understanding the electronic structure and
bonding of BaReH$_9$,
as schematically illustrated in Fig. \ref{levels}.
Specifically,
the d-like states of the H$_9$
shell have angular character about the Re site that favors bonding
with the Re d-states. This yields five bonding states, which are
$\sigma$ bonding combinations in the radial direction, and five
antibonding states, which have nodes between the Re and the H shell.
This provides an explanation for
the asymmetry of the valence and conduction bands,
since the antibonding combinations will be directed away from the
(ReH$_9)^{2-}$ clusters, leading to greater dispersion than for the
bonding combinations.

The strong bonding of the $d$-like combinations of H $s$ states on the H$_9$
shell and Re $d$ states leads to a large band gap, and shifts the bonding
states so that they are in the same energy region as the $p$-like
combinations of shell states, which are non-bonding with Re $d$ orbitals.
Thus, the 16 electron main peak in the valence DOS comes from
10 electrons in bonding Re $d$ - shell $d$ states, and 6 electrons
in shell $p$-like states. Considering the large bonding - antibonding splitting
of $\sim$ 8 eV as measured by the energy separation of the centers
of the valence and conduction Re $d$ DOS, there should be approximately
equal mixtures of Re $d$ and H $s$ character in both the bonding
and antibonding states. This would yield an electron count of
13 H 1$s$ electrons in the valence bands (including the $s$-like and $p$-like
shell states) and therefore a hydrogen valence of H$^{0.44-}$, in reasonable
accord with the estimate obtained from integrating the projected
DOS.

\section{Hypothetical Manganese Based Compounds}

The bonding mechanism discussed above is quite general in principle,
depending on a proper electron count
and the formation of strong bonds between $d$ orbitals and
appropriate combinations of H $s$ orbitals.
The $d$ orbitals of 3$d$ elements are much less extended than
those of 4$d$ and 5$d$ elements.
Thus, as is well known, 3$d$ compounds typically have smaller band widths
and reduced covalency with ligands than 4$d$ and 5$d$ analogues.
This might be used to provide
an explanation for why salts containing (MnH$_9)^{2-}$
have not been reported. Nonetheless, considering that the bonding in
BaReH$_9$ appears to be very strong, based on the large bonding - antibonding
splitting and the fact that the phase competes with the very stable hydride
BaH$_2$, it is of interest to study the Mn based analogue. We note
that the binding energy of MgH$_2$ is less than half of that of BaH$_2$,
and that the binding energy of BeH$_2$ is even lower, and so even with
reduced stability of the anionic cluster, salts may be formed with
appropriate cations.

In order to investigate the electronic structure of hypothetical
BaMnH$_9$, we reduced the $a$ and $c$ lattice parameters of
the Re compound based on the difference in the covalent radii
of Re and Mn (0.11\AA), to obtain $a$=5.067\AA, and $c$=8.883\AA.
This corresponds to a volume reduction of 12.5\%
relative to BaReH$_9$. Using these lattice parameters, we
relaxed the H positions in the unit cell. The resulting
H coordinates and bond lengths are give in Table \ref{struct-tab}.
Remarkably, the relaxed structure maintains a cage of nine
H atoms at approximately the same distance from Mn, and with
short H-H distances. 

\begin{figure}
\vskip 0.4 cm
\includegraphics[width=2.4in,angle=270]{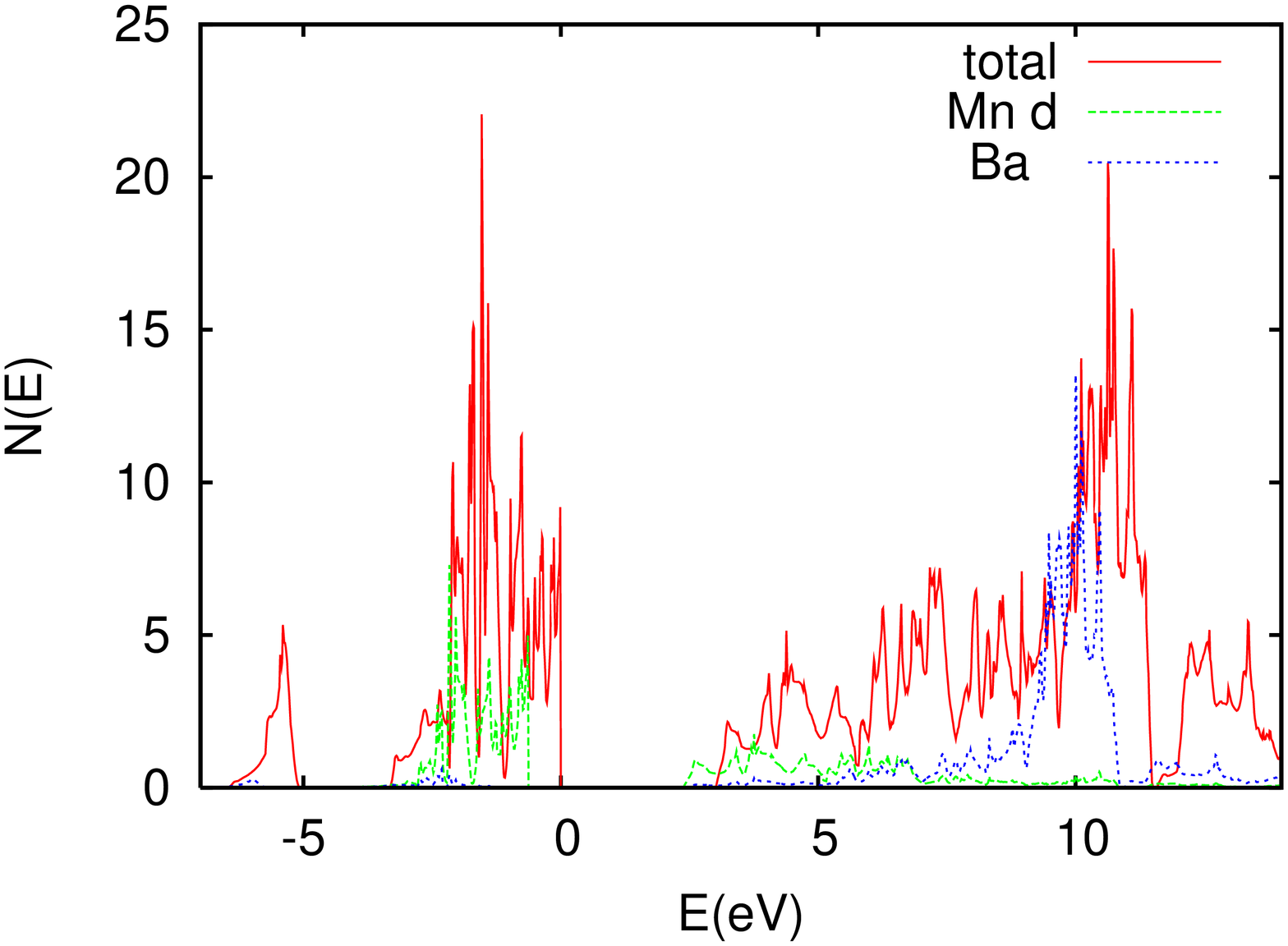}
\includegraphics[width=2.4in,angle=270]{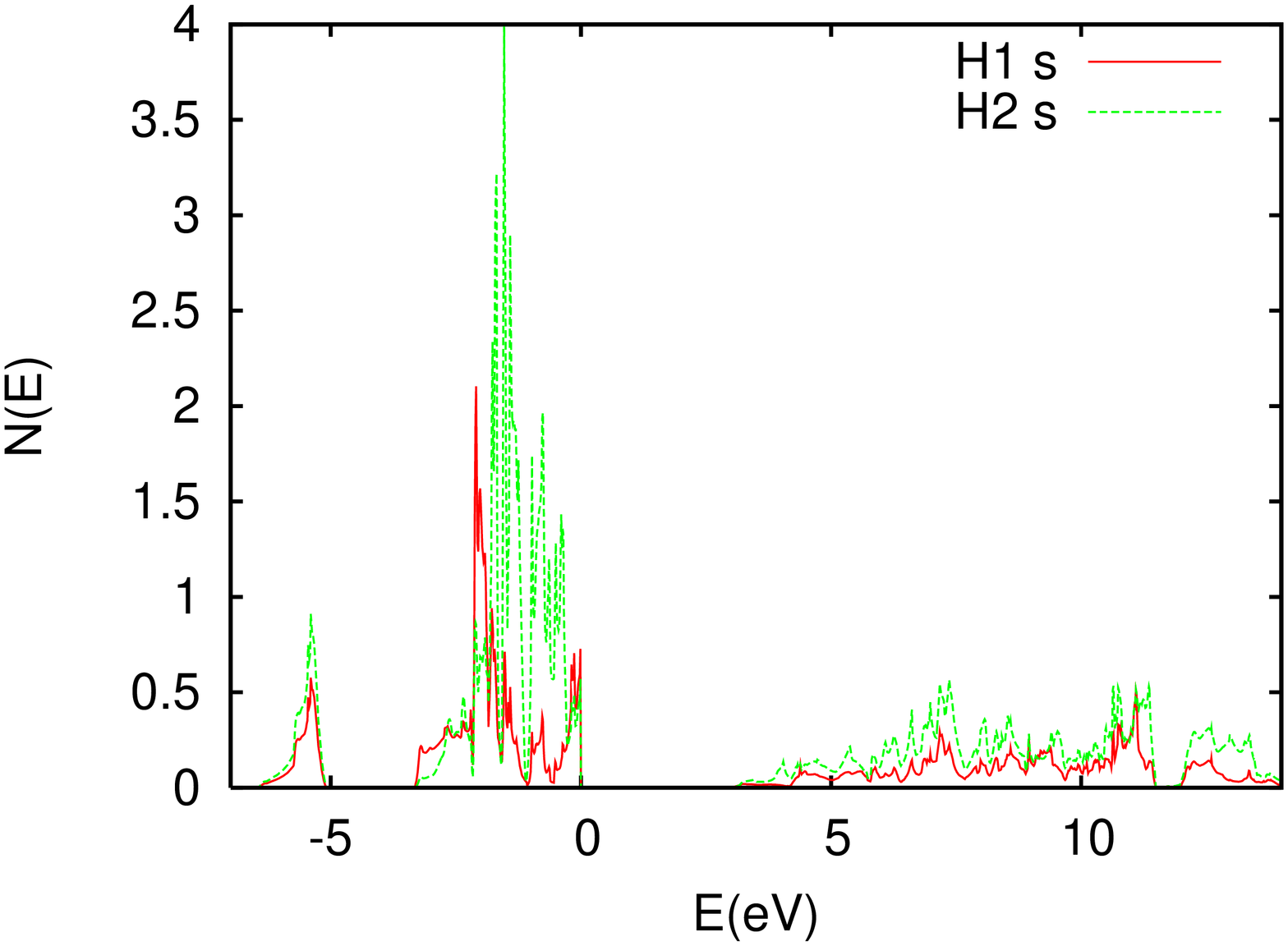}
\vskip 0.4 cm
\caption{(color online) Calculated electronic density of states
(DOS) of hypothetical BaMnH$_9$ on a per formula unit basis,
as in Fig. \ref{dos}. Note that the H LAPW sphere radius
for BaMnH$_9$ was 1.05 $a_0$.
}
\label{dos-mn} \end{figure}

As may be seen from the DOS and projections
(Fig. \ref{dos-mn}), the electronic structure is qualitatively
similar to BaReH$_9$, although the detailed peak structure of the
main valence band DOS differs and the split-off H state moves
to lower energy. This larger splitting between the shell $s$-like
state and the shell $p$-like states is what is expected based on
the smaller H-H distances in the more compact H$_9$ cage of the
Mn compound, since this will yield larger hopping matrix elements
and therefore stronger shell structure.
As in the Re compound, integration of the H $s$ projection of the
DOS, normalized by the amount of charge within a 1.05 $a_0$ sphere
for a neutral non-spin-polarized H atom in the LDA, yields a
hydrogen valence, H$^{0.5-}$. Furthermore, although the band gap
of BaMnH$_9$ is reduced to 3.0 eV, the bonding - antibonding splitting
as measured by the centers of the Mn projection of the DOS is
nearly the same as in the Re compound.

The implication of the above results is that salts containing
(MnH$_9)^{2-}$ may be realizable. Synthesis may, however, be
difficult because of competition from stable hydride phases.
As noted, MgH$_2$ is less stable than BaH$_2$. However, MgH$_2$ still
has a binding energy of 75 KJ/mol H$_2$, and furthermore Mg
forms other stable hydrides with Mn, such as Mg$_3$MnH$_7$. \cite{bortz,gupta}

The LDA often overbinds solids. Nonetheless, it is of interest to
compare the heats of formation of BaReH$_9$ and hypothetical
BaMnH$_9$. We calculated these values by subtracting calculated total
energies of the elemental metals and the H$_2$ molecule from the
total energy of the compounds. For Mn, we used the fcc structure
and made a correction of 5 mRy per atom, which comes from energy 
difference between the fcc structure and the ground state non-collinear
antiferromagnetic $\alpha$-Mn, as obtained by Hobbs and Hafner.
\cite{hobbs}
We made no correction for zero point energy.

With the LDA value of the energy of H$_2$, as obtained
in a supercell calculation, $E({\rm H}_2)$ = -2.294 Ry,
the resulting formation energies are -99 KJ/mol H$_2$ for
BaReH$_9$ and -86 KJ/mol H$_2$ for hypothetical BaMnH$_9$.
The main uncertainties are (1) the crystal structure of BaMnH$_9$,
which if different from that assumed would lead to a higher
stability, (2) the neglected zero point motion, which typically
affects the cohesive energies by less than 20 KJ/mol H$_2$ and
here would likely destabilize the compounds as the H-metal and
H-H bonds are short, suggesting stiff phonons, and 
(3) the use of the LDA, which based on calculations for
other metal hydrides,
\cite{smithson,halilov,miwa}
overestimates stability typically
by $\sim$ 20 KJ/ mol H$_2$, though this varies from compound to compound.
In any case, based on these results, we conclude that BaReH$_9$ and
hypothetical BaMnH$_9$ have similar stabilities, assuming that the
decomposition is to similar products, {\em e.g.} BaH$_2$ and Re/Mn metal.

\section{summary and conclusions}

Electronic structure calculations show that the high coordination in
BaReH$_9$ is due to bonding between Re 5$d$ states
and $d$-like states formed from combinations
of H $s$ orbitals in the H$_9$ cage. This explains the structure of the
material, its short bond lengths and other physical properties, such
as the high band gap. Similar bonding is found in hypothetical BaMnH$_9$
and in fact both compounds are found to have similar cohesive energies.
While BaMnH$_9$ may not be the most favorable target due to the
stability of BaH$_2$ as a competing phase, our results suggest that
synthesis of (MnH$_9)^{2-}$ salts may be possible and should be
attempted. We note that the chemistry of Mn VII is different
than that of Re VII. For example, KMnO$_4$ is highly oxidizing
in aqueous solution, while
KReO$_4$ is not. Thus, while it may be possible to synthesize (MnH$_9)^{2-}$
salts, the solvents and synthesis path needed may be different from that
used in the (ReH$_9)^{2-}$ salts.
Light element salts of (MnH$_9)^{2-}$ would have very high hydrogen
contents, in excess of 10 weight \% in the cases of Li, Mg or Be.

\acknowledgments

DJS thanks the University of Paris-Sud for their hospitality,
which made this work possible.
Crystal structures were plotted using the XCrySDEN program. \cite{xcrysden}
Work at Oak Ridge National Laboratory is supported by the U.S. Department
of Energy.
We thank the Institut du Developpement et des Ressources en
Informatique Scientifique (IDRIS) for a grant of computer time.

\end{document}